%% file: SCAPES.tex
\documentclass[twoside,a4paper]{article}
\input{preamble/auxiliar.tex}
\input{preamble/authors.tex}

\begin{document}
\ifpdf 
  \DeclareGraphicsExtensions{.png,.jpg,.pdf}
\else  
  \DeclareGraphicsExtensions{.eps}
\fi

\maketitle

\begin{abstract}
As generative audio models grow in complexity, the computational and ecological costs of synthesizing everyday sounds have become increasingly prohibitive, often requiring industrial-scale resources and massive datasets. In this paper, we present SCAPES: a Semantically Conditioned Autoregressive Prior for Environmental Sounds. SCAPES is a lightweight, resource-efficient generative model designed to synthesize high-fidelity environmental textures through high-level semantic control. By operating on the continuous latent manifold of a neural audio codec, our approach bypasses the rigid structural constraints inherent to discrete tokenization. We propose a segmentation strategy that decomposes audio into overlapping segments, enabling a Continuous Normalizing Flow (CNF) to model the evolution of  latent trajectories using Flow Matching. Our experiments demonstrate that a 36-million parameter instance of SCAPES can be trained on limited, uncurated datasets using a single consumer-grade GPU. Notably, convergence is achieved after training for approximately twice the source audio duration, yielding high-fidelity outputs with robust long-term stability and semantic consistency. Furthermore, we showcase the model's capacity for smooth semantic interpolation, providing a flexible and accessible tool for open research and creative sound design. Code, pretrained weights, audio examples, and an interactive demo are publicly available on our project page \href{https://cordutie.github.io/projects/scapes.html}{\texttt{cordutie.github.io/projects/scapes.html}}.
\end{abstract}

\section{Introduction}
\label{sec:intro}
The concept of ``everyday listening'' emphasizes the human experience of hearing events in the world rather than the abstract properties of the sounds themselves \cite{gaver1993_what}. While traditional ``musical listening'' focuses on perceptual dimensions like pitch, loudness, and timbre, everyday listening is primarily concerned with the physical sources, materials, and the underlying semantic meaning of the interactions that produce the audio.

When viewed through this ecological lens, the acoustic domain reveals itself as a vastly diverse field of environmental sounds. This broad spectrum encompasses everything from discrete, structured interactions, such as keyboard strokes, applause, or a helicopter rotor, to continuous, stochastic textures, like rushing rivers, crackling bonfires, wind blowing, or heavy rain.

With the rapid advancement of generative audio models in recent years, there has been a growing research focus on \textit{controllability} \cite{DDSP,AFTER,sketch2audio,sketch2sound,musgu}, defined as the ability to intuitively guide the generative process. Nevertheless, designing intuitive control mechanisms for environmental sounds remains a significant challenge, owing to their diversity and lack of inherent structure. In the context of generative audio models, user-controllable features can be broadly categorized as follows:
\begin{enumerate}
    \item \textbf{Signal properties:} Purely signal-based metrics, such as spectral coefficients, zero-crossing rate, and signal energy.
    \item \textbf{Perceptual features:} Properties derived from psychoacoustic models, including pitch, loudness, and perceptual summary statistics.
    \item \textbf{Data-driven representations:} Abstract features learned from the data distribution, such as unconditioned latent spaces.
    \item \textbf{Source properties:} Low-level physical parameters of the sound-generating mechanism, such as string length or material density.
    \item \textbf{Semantic features:} High-level, human-intuitive descriptors, such as text prompts or sound event labels.
\end{enumerate}

In the realm of open, lightweight, and controllable sound synthesis, various models have successfully leveraged a subset of these modalities. Pioneering physical modeling approaches \cite{KarplusStrong, OBrien2002} are fundamentally driven by \textit{source properties}. Texture synthesis frameworks, such as those introduced by McDermott and Simoncelli \cite{mcdermott2011, mcdermott2013} alongside TexStat \cite{texstat}, generate audio governed by \textit{perceptual} summary statistics. Differentiable digital signal processing (DDSP) techniques \cite{DDSP, noisebandnets} typically combine \textit{signal} and \textit{perceptual} features to guide the generation process. Additionally, alternative texture synthesis models like SynTex \cite{syntex} rely on varying combinations of \textit{source} and \textit{signal} properties. Finally, modern neural audio models based on Variational Autoencoders, including \cite{aaron} and other widely recognized architectures such as RAVE \cite{RAVE} and AFTER \cite{AFTER}, synthesize sound by navigating \textit{data-driven representations} defined inherently by the model architecture.

Since everyday listening is fundamentally anchored in high-level semantics, \textit{semantic features} offer the most intuitive modality for controlling everyday sound generation. However, their inherent coarseness lacks the fine-grained structural information needed to directly synthesize complex waveforms. Mapping such sparse, high-level descriptors to dense audio signals is a highly under-constrained problem; consequently, successful approaches have largely depended on leveraging immense datasets to learn these complex distributions. As a result, semantic control is currently dominated by large-scale text-to-audio frameworks \cite{woosh, makeanaudio, audioldm, sketch2audio}, with a growing emphasis on closed-source, proprietary systems such as Meta's Audiobox \cite{audiobox}, Adobe's Sketch2Sound \cite{sketch2sound}, and Nvidia's Fugatto \cite{fugatto}. Ironically, this reliance on scale implies that synthesizing even simple, natural environmental sounds requires deploying highly resource-intensive models, an approach that carries a substantial ecological footprint.

To address this, we present SCAPES: a Semantically Conditioned Autoregressive Prior for Environmental Sounds. SCAPES is a generative model trained without manual annotations, designed for high-fidelity, semantically controllable stereo sound synthesis with minimal data and computational cost. It overcomes the coarseness of traditional semantic descriptors by injecting high-level embeddings at fine temporal resolutions, enabling continuous, granular control over generated sounds while staying within the learned semantic manifold.

At its core, the SCAPES architecture is a small-scale Continuous Normalizing Flow (CNF) \cite{cnf} that serves as a conditional prior over continuous EnCodec \cite{encodec} latent representations. Unlike traditional autoregressive architectures that predict sequences token-by-token, the SCAPES architecture is designed to synthesize short audio segments in a single pass. The SCAPES model, however, applies this architecture sequentially over time, conditioning each segment on a memory buffer of previous latent states and high-level semantic embeddings from CLAP \cite{clap_microsoft, clap_laion}. Consequently, the generative process is autoregressive at the segment level, while the underlying architecture remains a non-autoregressive, conditional vector field.

The vector field governing the flow is parameterized by a Transformer \cite{transformer} and optimized via Flow Matching \cite{flow-matching}. Recently, flow-based models have gained significant traction in audio synthesis due to their robust generative capabilities, stable training dynamics, and capacity to model complex continuous distributions without the information loss inherent to discrete, token-based approaches. While these frameworks have been primarily adopted by large-scale, latency-critical text-to-audio systems, SCAPES demonstrates their effectiveness within a lightweight, resource-efficient regime. Furthermore, although not explored in the present work, inference speed can be further accelerated through rectification and distillation processes \cite{rectifiedflows}.

This paper is structured as follows: Section~\ref{sec:background} establishes the mathematical and methodological background for the proposed architecture and evaluation metrics. Section~\ref{sec:model_design} details the SCAPES model design, encompassing data representation, architecture, and inference strategies. Section~\ref{sec:training_and_evaluation} presents a comprehensive case study, detailing the dataset, training efficiency, and an evaluation of generation quality, semantic adherence, and long-term stability. Section~\ref{sec:discussion} provides a detailed discussion of the model's high degree of controllability, its capacity for novel semantic interpolation, and its current limitations. Finally, Section~\ref{sec:conclusions} concludes the paper with a summary of contributions and directions for future work.

\section{Background}\label{sec:background}
This section establishes the foundational mathematical frameworks and methodological concepts underlying the proposed architecture, alongside the specific techniques utilized for generation quality assessment.

\subsection{Neural Audio Codecs}\label{subsec:codec}
Neural Audio Codecs (NACs) are deep learning architectures designed to compress high-fidelity audio into low-dimensional latent representations for efficient processing, before reconstructing the signal back to the time domain. Recent advancements in NACs generally fall into two paradigms. The first employs a discrete bottleneck, utilizing processes like Residual Vector Quantization (RVQ) to compress latents into discrete token vocabularies, as seen in EnCodec \cite{encodec} and DAC \cite{dac}. The second paradigm seeks to move away from strictly discrete bottlenecks, focusing either on purely continuous representations, as proposed by Music2Latent \cite{m2l}, or hybrid continuous-discrete architectures like CodiCodec \cite{codicodec}. Because the proposed model builds upon the 48\,kHz stereo, non-causal variant of EnCodec, we briefly detail its specific formulation here.

In the 48\,kHz EnCodec architecture, the input waveform is first normalized, yielding a single scalar factor representing the audio scale. This scale factor is explicitly retained, as it is essential for restoring the signal's original dynamic range upon reconstruction. The normalized audio is then processed through a stack of convolutional encoder layers, progressively downsampling the temporal sequence into a compact 128-dimensional continuous latent space. It is critical to note that the convolutional operations in this specific implementation are \textit{non-causal}; they rely on a receptive field that requires access to both past and future context to accurately encode the current frame. Typically, these 128-dimensional vectors are subsequently passed through an RVQ module to construct a discrete token sequence. However, when utilizing the codec purely for its continuous representation, this quantization step is bypassed. The unquantized 128-dimensional latents, augmented by their corresponding scale factors, can be fed directly into the non-causal convolutional decoder to be expanded back into a high-fidelity time-domain waveform.

\subsection{Continuous Normalizing Flows and Flow Matching}\label{subsec:cnf_and_fm}
Continuous Normalizing Flows (CNFs) \cite{cnf} are a class of generative models built upon the framework of Neural Ordinary Differential Equations (Neural ODEs). In this paradigm, rather than relying on a discrete composition of functions as in traditional deep networks, the transformation between distributions is modeled as a continuous-time flow. More precisely, given a time-dependent vector field $F: \mathbb{R}^d \times [0,1] \rightarrow \mathbb{R}^d$, the generative mapping is governed by the dynamics of the Cauchy initial value problem:
\begin{equation}\label{eq:dynamics}
\left\{
\begin{array}{rl}
\dfrac{dx}{dt} &= F(x(t), t)\\
x(0)          &= x_0
\end{array}
\right.
\end{equation}
where the initial condition $x_0 \in \mathbb{R}^d$ serves as the model's input and is typically sampled from a tractable base distribution (e.g., a standard Gaussian). The generated output is then obtained by evaluating the ODE solution at the final evolution parameter $t=1$, yielding $x_1=x(1)\in \mathbb{R}^d$.

The central learning objective in a CNF is to approximate the true vector field using a parameterized neural network $F_\theta$. Flow Matching \cite{flow-matching} provides an efficient framework to achieve this, and its mechanism is briefly explained here. First, for any paired sample $(x_0, x_1)$ drawn from the base and target data distributions respectively, one must construct a \textit{conditional vector field} $F(x, t \,|\, x_1)$ that explicitly guides the dynamics from $x_0$ to $x_1$. Specifically, integrating this conditional vector field through \eqref{eq:dynamics} reveals a deterministic \textit{conditional path} $x_t = x(t \,|\, x_0, x_1)$ connecting the two states. It can be shown that the intractable target vector field, which pushes the entire base distribution to the true data distribution, can be learned by regressing against these conditional fields. This is achieved by optimizing the \textit{Conditional Flow Matching} (CFM) objective:
\begin{equation}
\mathcal{L}_{\text{CFM}}(\theta) = \mathbb{E}_{t, x_0, x_1} \left\| F_\theta(x_t, t) - F(x_t, t \,|\, x_1) \right\|^2,
\end{equation}
where $t \sim \mathcal{U}(0,1)$. Minimizing this objective guarantees that the neural vector field $F_\theta$ marginalizes over the conditional vector fields, lifting their transport properties globally to recover the complete data distribution (see \cite[Theorem 2]{flow-matching}).

\subsection{Generation Quality Assessment}\label{subsec:fad_and_kad}
The Fréchet Audio Distance (FAD) \cite{fad} is a standard metric for evaluating generative audio models, measuring the dissimilarity between the mean and covariance of different embedding distributions. However, FAD inherently assumes a multivariate Gaussian distribution, an assumption that often fails for complex data. Moreover, it suffers from finite-sample bias and computationally expensive covariance operations in high dimensions \cite{gui2024adapting}. These limitations can be overcome with the Kernel Audio Distance (KAD) \cite{kad}. Based on the Maximum Mean Discrepancy (MMD) framework, KAD is a non-parametric metric that compares distributions in a kernel-induced feature space, capturing complex non-linear statistics without imposing restrictive distributional assumptions.

Both metrics are fundamentally tied to the chosen representation space, and many representations are valid in this context. To ensure a comprehensive evaluation, we use three embedding spaces. First, while acknowledging its statistical limitations, FAD is computed using standard VGGish embeddings \cite{vggish} to maintain baseline comparability with existing literature. Second, KAD is computed using CLAP embeddings \cite{clap_microsoft, clap_laion} to provide a robust assessment of high-level semantic information. Finally, KAD is applied to perceptual summary statistics derived from TexStat \cite{texstat}, based on the McDermott and Simoncelli texture synthesis framework \cite{mcdermott2011, mcdermott2013}. By encoding the distributions of cochlear subband envelopes, modulation energies, and cross-channel correlations, TexStat provides a targeted evaluation of the perceptual quality of environmental texture sounds.

\section{Model Design}\label{sec:model_design}
SCAPES is designed to synthesize high-fidelity environmental so-unds by modeling the distribution of localized latent segments, referred to here as ``atoms.'' Each atom is generated by a conditional vector field that accounts for a preceding sequence of atoms and a high-level semantic context embedding. This specific data representation is foundational to the model's architecture and performance. In this section, we detail the formulation of the atomic representation and the rationale behind its design. Subsequently, we describe the architectural framework, focusing on the conditioning mechanisms and the iterative inference strategy. For the sake of generality, we describe all variables abstractly here. For concrete numeric examples see Section~\ref{sec:training_and_evaluation}.

\subsection{Data Structure: Atomic Segmentation and Semantic Annotations}\label{subsec:dataprep}
SCAPES is agnostic to the specific choice of Neural Audio Codec, provided it utilizes continuous latent representations. This design avoids the ``topological shattering'' inherent to discrete tokenization, where the acoustic manifold is fragmented into a disjointed set of indices. By bypassing these rigid structural constraints, the model preserves the natural continuity of the audio signal. For this work, we validate our framework using the continuous latent space of the 48\,kHz EnCodec non-causal variant.

\begin{figure}[ht]
    \centering
    \includegraphics[width=\columnwidth]{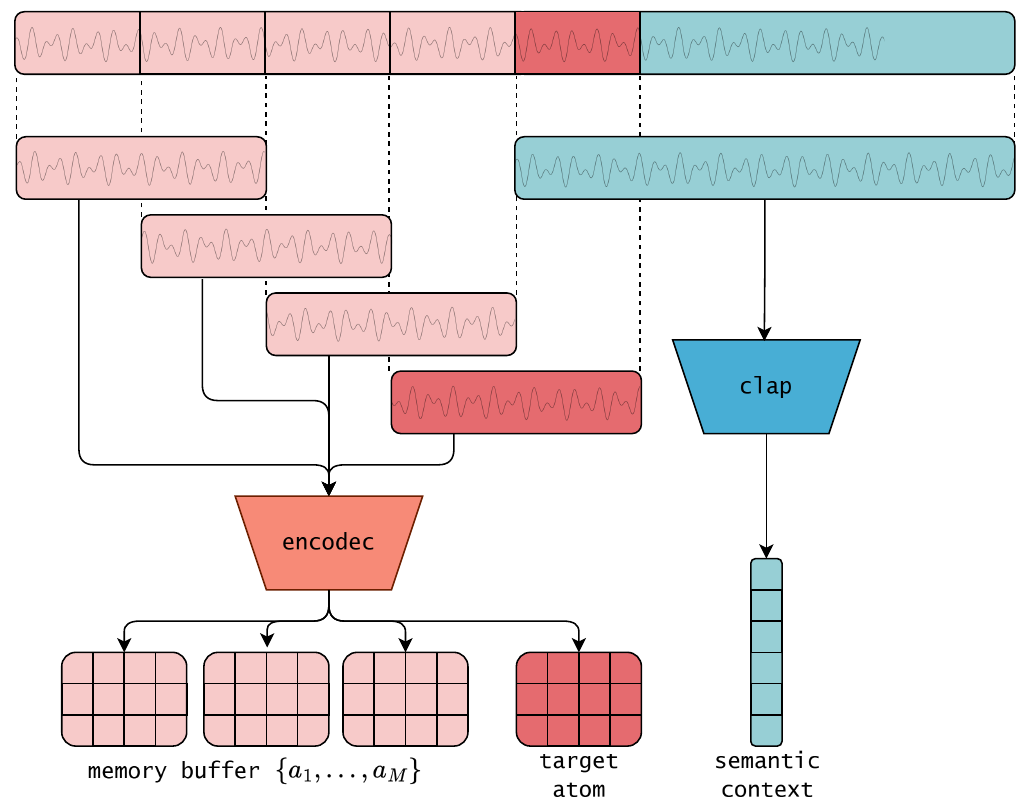}
    \caption{Data representation and annotations. Source audio (top row waveform) is decomposed into overlapping segments (pink and red waveforms) and encoded with EnCodec into \textit{atoms} (pink and red matrices). Each atom is annotated with a semantic embedding (cyan vector) computed from a context window (cyan waveform), extended through randomized repetitions to satisfy CLAP's minimum input length. During training, the model is tasked to predict the target atom (red matrix) from the preceding atoms in the memory buffer and its corresponding semantic embedding.}
    \label{fig:dataprep}
\end{figure}



Because EnCodec utilizes non-causal convolutions and relies on a global normalization scale factor, naively concatenating independently generated latents produces severe boundary artifacts and phase inconsistencies. To resolve these conflicting receptive fields, we propose an \emph{atomic} segmentation strategy inspired by Saint-Arnaud's pioneering work on sound textures \cite{saintarnaud1995}. We split the continuous audio into overlapping segments and encode each with EnCodec, producing a 128-dimensional latent sequence and a single global scale factor per segment. We combine these into an \textit{atom} $a \in \mathbb{R}^{(128+1) \times N},$ where $N$ is the number of temporal frames. The first 128 dimensions hold the latent features, and the last dimension contains the scale factor, broadcasted along the temporal axis. This unified representation ensures that the latent trajectories and local normalization can be learned together.

While atoms act as a continuous analogue to discrete tokens, they lack the explicit boundary constraints of a fixed vocabulary. In token-based systems, the decoder learns to resolve transitions between discrete indices; in contrast, our atoms are generated independently. To provide contextual continuity, we introduce substantial temporal overlap between successive segments, encouraging the model to capture the joint structure of the sound through redundant acoustic information. This is not trivial because identical acoustic content can correspond to different latent representations, due to each atom's local normalization and non-causal receptive field. Our experiments show that a Transformer-based architecture effectively reconciles these variations, allowing the model to maintain coherence throughout generation, as further detailed in Section~\ref{sec:training_and_evaluation}.

Each atom is further paired with a semantic context embedding $c$, derived from a pre-trained CLAP model \cite{clap_microsoft,clap_laion}. While the CLAP implementation utilized requires 7-second audio windows to produce high-level embeddings, we aim for a finer level of annotation. To achieve this, we extract a context window containing the target atom's audio, typically shorter than 2 seconds, and extend it to the required 7-second duration via overlapped and added repetition of internal segments of random sizes. This approach generates a continuous, textural extension of the source audio that deliberately avoids introducing artificial rhythmic periodicities, preserves the original timbral characteristics, and provides a robust, localized semantic descriptor.

In summary, this data structure ensures that during training each target atom $a_{\text{target}}$ can be conditioned on a preceding sequence of atoms $\{a_1, \dots, a_M\}$, which serve as the model's temporal memory, along with a corresponding semantic context embedding $c$. The resulting data pipeline is illustrated in Figure~\ref{fig:dataprep}.

\subsection{Architecture and Flow Matching}\label{subsec:architecture}
The SCAPES architecture consists of a Continuous Normalizing Flow (CNF) where the time-dependent vector field $F_\theta$ is parameterized by a Transformer. The primary objective of the model is to learn the conditional distribution of a target atom $a_{\text{target}}$ given a sequence of preceding atoms $\{a_1, \dots, a_M\}$ and a semantic context embedding $c$. This localized generation task allows for a significantly reduced parameter count while maintaining high fidelity, as the model is only responsible for synthesizing short-duration, continuous latent representations rather than long-form waveforms.

The Transformer $F_\theta(x, t \,|\, \{a_1, \dots, a_M\}, c)$ processes its inputs through three integrated mechanisms. Within each atom, \textit{self-attention} sub-layers model the local intra-atomic temporal structure using Rotary Positional Embeddings (RoPE) to encode frame-level positions. To ensure coherence across atoms, \textit{cross-attention} sub-layers allow the model to selectively retrieve relevant acoustic context from a memory buffer composed of the $M$ preceding atoms, augmented with a learned atom-position embedding. Finally, global conditioning is incorporated via \textit{Adaptive Layer Normalization Zero} (AdaLN-Zero) \cite{adalnzero}, where a dedicated MLP modulates the network's hidden states based on the flow-time $t$ and the semantic context $c$.

Training is conducted using the Conditional Flow Matching objective, applied independently to the 128 latent dimensions and the scale dimension. We adopt the Optimal Transport (OT) conditional vector field \cite{flow-matching}. In this framework, the conditional vector field and paths correspond to
\begin{align}
F(x, t \,|\, x_1) &= \frac{x_1 - (1 - \sigma_{\min})x}{1 - (1 - \sigma_{\min})t},\\
x(t\,|\,x_0,x_1)  &= (1 - (1 - \sigma_{\min})t)x_0 + tx_1,
\end{align}
where $x_1$ represents the target atom data, $x_0 \sim \mathcal{N}(0, I)$ is the noise from the base distribution, and $\sigma_{\min}$ is a small variational regularizer used to ensure stability near $t=1$. This formulation reveals conditional paths corresponding to straight lines and reduces the CFM objective to
\begin{equation}
\mathcal{L}_{\text{CFM}}(\theta) = \mathbb{E}_{t, x_0, x_1} \left\| F_\theta(x_t, t) - (x_1 - (1 - \sigma_{\min})x_0) \right\|^2,
\end{equation}
where, for clarity, the notation suppresses the conditioning on previous atoms $\{a_1, \dots, a_M\}$ and semantic context $c$ in $F_\theta$.

\subsection{Inference and Continuous Synthesis}\label{subsec:inference}
\begin{figure}[h]
    \centering
    \includegraphics[width=\columnwidth]{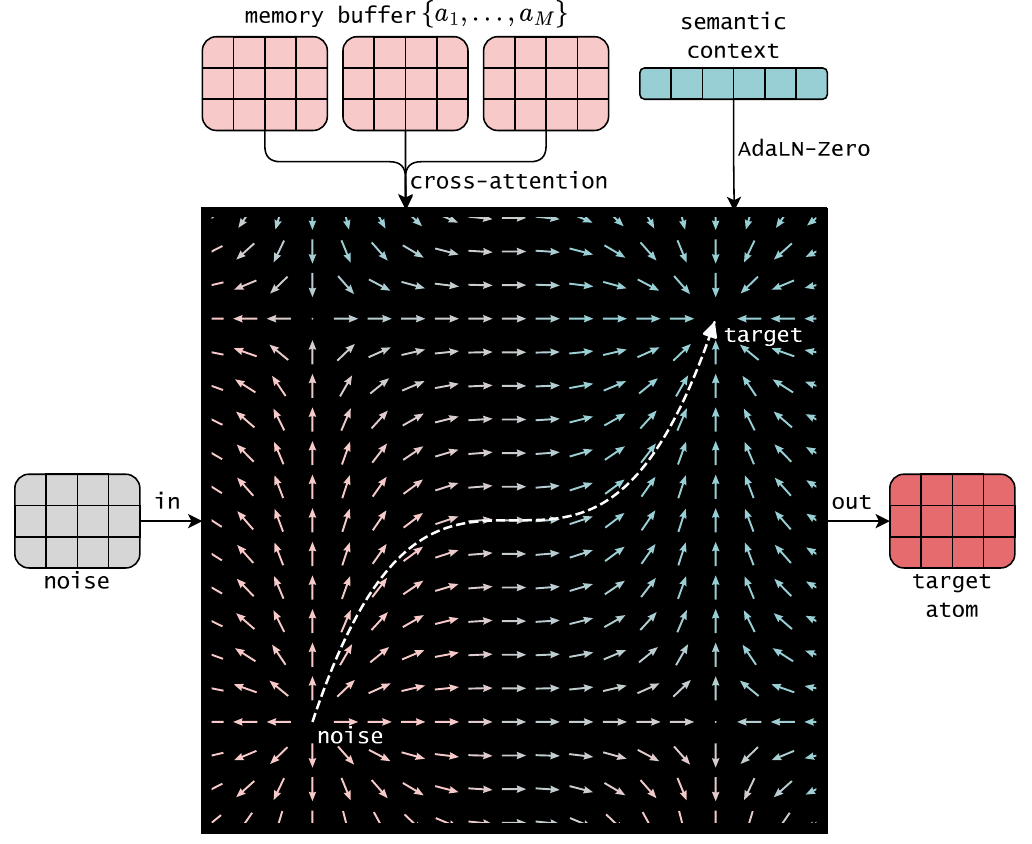}
    \caption{Inference architecture. The Transformer-based vector field $F_\theta$ is conditioned on past atomic memory via cross-attention and semantic context via AdaLN. An ODE solver integrates the field from $t=0$ to $t=1$ to generate the target atom.}
    \label{fig:arch}
\end{figure}

At inference time, the generative process is formulated as an iterative sequence of ODE initial value problems. Each atom is synthesized by solving the Cauchy problem \eqref{eq:dynamics} over the interval $t \in [0, 1]$ using a second-order Runge-Kutta method with a fixed number of function evaluations to balance synthesis speed and numerical precision. The vector field $F_\theta$ is conditioned on the semantic context $c$ and a sliding memory buffer $\{a_1, \dots, a_M\}$, with Gaussian noise $\mathcal{N}(0, I)$ serving as the initial condition $x_0$. This procedure is illustrated in Figure~\ref{fig:arch}. Once a new atom $x_1=a_{M+1}$ is generated, it is appended to the memory buffer while the oldest atom is discarded, maintaining a consistent temporal context for the subsequent step.

To assemble these atoms into a continuous stream, we utilize a modified Overlap-and-Add (OLA) procedure that builds upon the structural consistency enforced during training.
To achieve perceptually transparent transitions, we zero mask the overlapping portion of the new atom, reserving 0.02\,s to cross fade between the previous atom tail and the current atom head. This recursive process maintains phase continuity and enables the stable synthesis of coherent, long-form acoustic textures.

\section{Training and Evaluation}\label{sec:training_and_evaluation}
SCAPES has been evaluated across multiple hand-curated datasets with promising results. This section details a specific case study that highlights the model's data requirements, training efficiency in terms of both time and hardware resources, and the subsequent assessment of generation quality. To support open research and reproducibility, all code, data, pretrained weights, audio examples, and a demo are publicly available on our project page\footnote{\href{https://cordutie.github.io/projects/scapes.html}{\texttt{cordutie.github.io/projects/scapes.html}}.}.

\subsection{Dataset and Training Specifications}\label{subsec:train_specs}
The evaluation dataset was curated from Freesound \cite{freesound} without manual editing and comprises approximately 34 minutes of audio across 10 environmental categories. 
The sound categories include concert hall applause, crackling bonfires, bubbling water, forest ambiances, a helicopter rotor, keyboard strokes, an orchestra tuning, rain of varying intensities, rivers of different sizes, and wind textures. Following Section~\ref{subsec:dataprep}, audio was segmented into atoms of 33 EnCodec frames (0.22\,s) with a hop size of 15 frames (0.1\,s). This configuration leaves 3 frames of overlap on each side (totaling 0.02\,s for cross-fading) and achieves a precise control rate of 10\,Hz. Additionally, the memory buffer stores $M=5$ past atoms, providing 0.5\,s of non-redundant audio context.

For this case study, we utilized a SCAPES configuration consisting of approximately 36 million parameters. The model employs 6 transformer layers, each with a hidden dimension of 512, 8 attention heads, and a feed-forward dimension of 2048. Every layer comprises self-attention, cross-attention, and feed-forward sub-layers. Adaptive Layer Normalization (AdaLN) is applied before each, resulting in 18 modulated sub-layers across the network, followed by a final AdaLN module prior to the output projection.

Training was conducted using the Conditional Flow Matching objective for 120 epochs with a batch size of 32 on a single NVIDIA RTX 4090 GPU. The process proved highly efficient, with peak VRAM usage remaining under 8 GB and total training time of approximately one hour. These benchmarks demonstrate that SCAPES can be successfully trained on consumer-grade hardware in a fraction of the time required by massive, token-based models, significantly democratizing high-fidelity sound synthesis for open research. Furthermore, inference is highly efficient: solving the ODE with 16 steps achieves generation speeds twice that of real-time. Nevertheless, to prioritize acoustic fidelity, we employed 32 steps for all experiments detailed in this study.

\subsection{Evaluation}\label{subsec:evaluation}
In this subsection, we benchmark the generative performance of SCAPES across four complementary dimensions. First, we evaluate objective audio quality by comparing the resynthesis capabilities of SCAPES against RAVE \cite{RAVE}, a well-established real-time neural audio synthesis model. Second, we assess semantic adherence by computing FAD and KAD scores to verify that the model maintains class integrity throughout the generative process. Third, we investigate the long-term stability of the autoregressive chain, addressing the "drift" problem typical of iterative latent modeling. Finally, we provide some qualitative observations on semantic interpolation, exploring the model's ability to navigate the continuous latent manifold between disparate acoustic categories.

\subsubsection{Generation Quality and Resynthesis}\label{subsubsec:resynthesis}
Evaluating SCAPES presents a unique challenge, as its small scale, controllability, stochastic generation, and multi-class capabilities make finding a direct baseline difficult. To facilitate a controlled evaluation of audio quality only, we approximate a resynthesis task by conditioning the model on ground truth past audio and precomputed CLAP context embeddings. We selected a RAVE model as our benchmark due to its similarly small scale, high fidelity reconstruction, and shared capacity for modeling multiple sound categories within one architecture. We acknowledge that this comparison is inherently asymmetrical: RAVE learns its latent space from scratch for complete signal reconstruction, whereas SCAPES learns only the data distribution on top of a frozen NAC with massive hidden compute already baked in. Nevertheless, the practical computational differences for a researcher or artist are significant. In our experimental setup, SCAPES operated on less than half the VRAM demanded by RAVE and reached convergence roughly 90 times faster.

To quantify performance, we employed the metrics described in Section~\ref{subsec:fad_and_kad}, namely FAD (VGGish), KAD (CLAP), and KAD (TexStat). These scores were computed between the original audio and the corresponding resynthesized outputs produced by each model, and are summarized in Table~\ref{tab:generation_quality}. The results show that, although RAVE demonstrates strong signal reconstruction capabilities, SCAPES generally achieves superior distributional similarity.

We hypothesize that the relatively homogeneous nature of the dataset poses a challenge for RAVE, limiting its ability to generalize effectively. In particular, we observe that RAVE struggles to accurately reconstruct highly noisy sounds. In contrast, SCAPES exhibits more robust performance across such cases. This suggests that the flow-matching objective is effective at capturing the underlying statistical texture of the environment, even in the absence of an explicit reconstruction objective.

\begin{table}[h]
    \centering
    \footnotesize
    \caption{Quantitative comparison of generation quality across various metrics and models. The best results for each metric (lower is better) are highlighted in \textbf{bold}.} 
    \setlength{\tabcolsep}{0pt}
    \begin{tabularx}{\columnwidth}{
    >{\raggedright\arraybackslash}X
    >{\centering\arraybackslash}X
    >{\centering\arraybackslash}X
    >{\centering\arraybackslash}X
    >{\centering\arraybackslash}X
    >{\centering\arraybackslash}X
    >{\centering\arraybackslash}X
    }
    \toprule
    & \multicolumn{2}{c}{\textbf{FAD} (VGGish)} & \multicolumn{2}{c}{\textbf{KAD} (CLAP)} & \multicolumn{2}{c}{\textbf{KAD} (TexStat)} \\
    \cmidrule(lr){2-3} \cmidrule(lr){4-5} \cmidrule(l){6-7}
    \textbf{Source} & SCAPES & RAVE & SCAPES & RAVE & SCAPES & RAVE \\
    \midrule
    Applause   & \textbf{4.35}  & 39.06          & \textbf{33.16} & 47.77          & \textbf{38.1}  & 65.32          \\
    Bonfire    & 3.66           & \textbf{3.65}  & 49.37          & \textbf{37.64} & \textbf{13.28} & 15.45          \\
    Bubbles    & \textbf{5.03}  & 21.05          & \textbf{31.89} & 47.82          & \textbf{10.9}  & 37.87          \\
    Forests    & 11.05          & \textbf{3.49}  & 61.71          & \textbf{52.56} & 37.28          & \textbf{16.38} \\
    Helicopter & \textbf{6.9}   & 8.0            & \textbf{22.65} & 27.8           & \textbf{11.17} & 28.88          \\
    Keyboard   & \textbf{1.05}  & 1.51           & \textbf{55.66} & 85.68          & \textbf{19.73} & 29.47          \\
    Orchestra  & \textbf{5.14}  & 11.42          & \textbf{17.91} & 37.62          & \textbf{3.14}  & 9.8            \\
    Rain       & \textbf{5.67}  & 13.38          & \textbf{66.85} & 92.02          & \textbf{15.4}  & 57.16          \\
    Rivers     & \textbf{10.17} & 23.16          & \textbf{76.49} & 85.14          & \textbf{39.33} & 58.86          \\
    Wind       & \textbf{18.69} & 23.6           & \textbf{30.36} & 37.25          & 23.71          & \textbf{7.19} \\
    \bottomrule
    \end{tabularx}

    \label{tab:generation_quality}
\end{table}

\subsubsection{Semantic Conditioning and Class Adherence}\label{subsubsec:semantic_adherence}
To evaluate the model's adherence to semantic conditioning, we performed fully autoregressive generation for each of the 10 acoustic categories. Each generation was initialized with a "cold start" (zero-initialized memory buffer) and conditioned on the CLAP embeddings derived from the training set. The resulting synthetic audio was then benchmarked against the original dataset using the distributional similarity metrics defined in Section~\ref{subsec:fad_and_kad}.

\begin{figure*}[h]
\centering
\hspace{-2mm}
\includegraphics[width=0.365\textwidth]{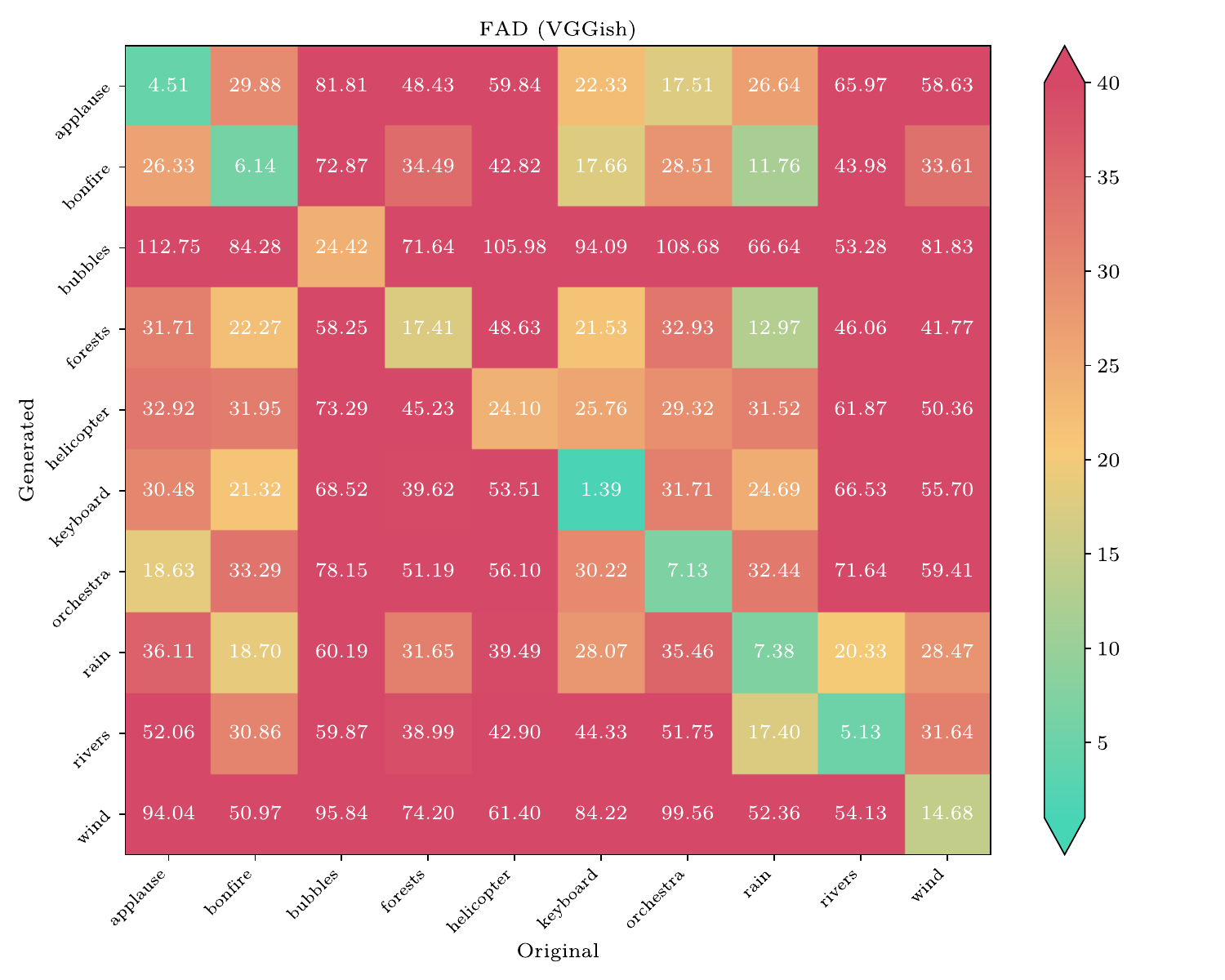}
\hspace{-5mm}
\includegraphics[width=0.30466\textwidth]{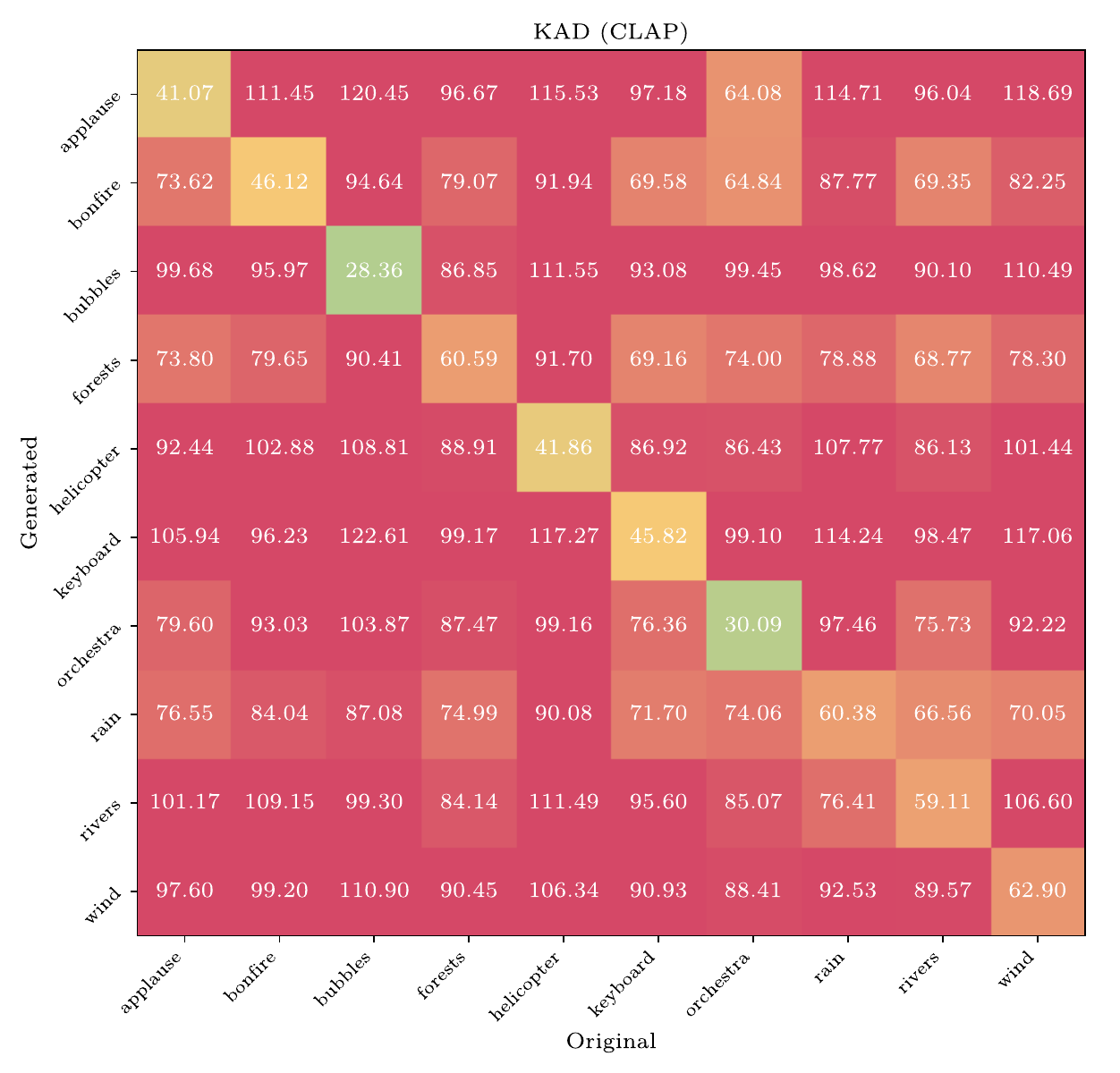} 
\hspace{-3mm}
\includegraphics[width=0.365\textwidth]{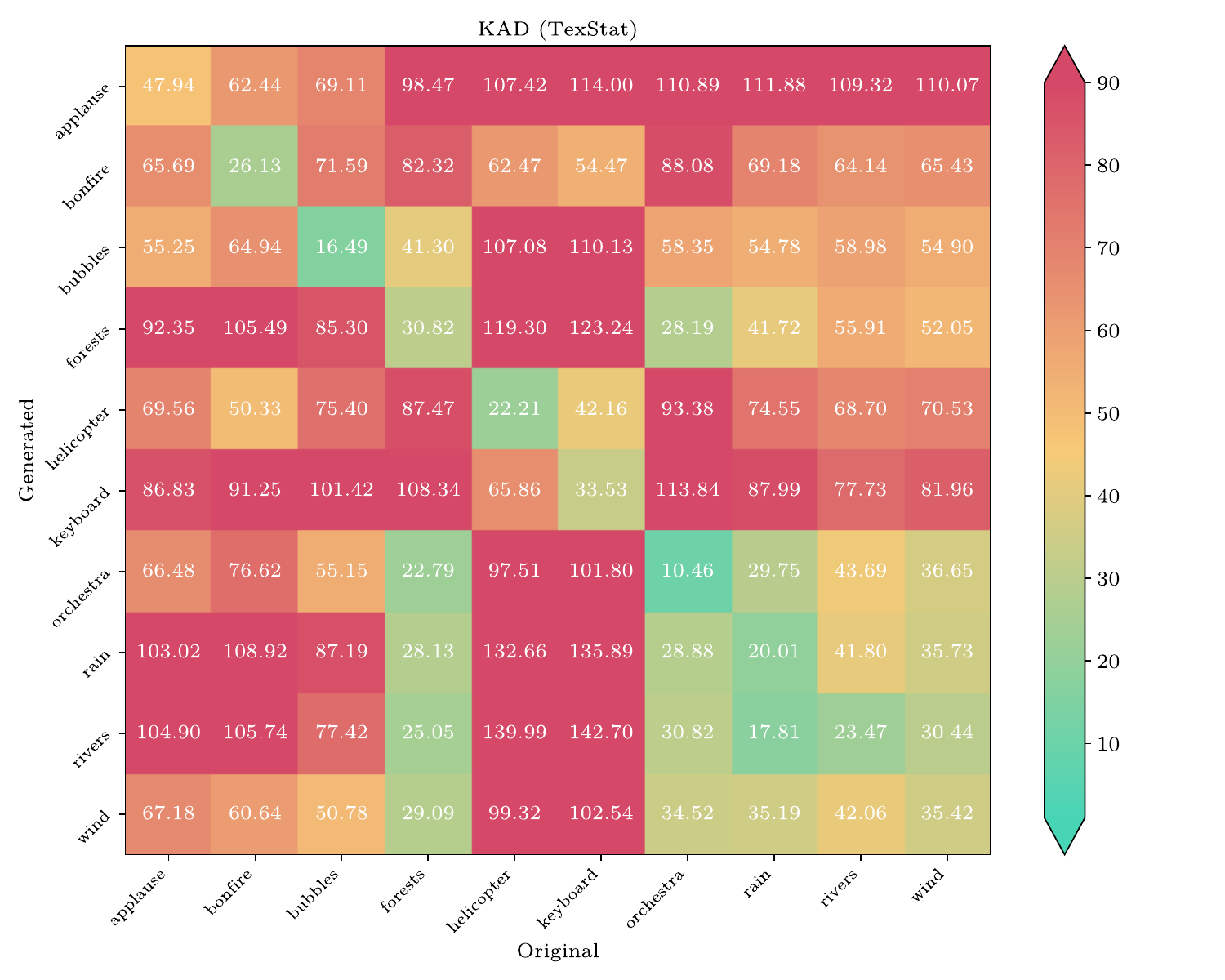}
\caption{Class-wise distance matrices for generated versus original audio. Lower values indicate greater acoustic similarity. From left to right: FAD (VGGish), KAD (CLAP), and KAD (TexStat), latter two sharing a colorbar. FAD and KAD are not directly comparable. The KAD scores, however, share the same scale due to the use of a common scale invariant kernel, although their interpretations differ.}
\label{fig:confusion_matrices}
\end{figure*}

Figure~\ref{fig:confusion_matrices} illustrates the resulting similarity matrices across the three evaluation frameworks. In all cases, the matrices exhibit a prominent diagonal structure, demonstrating that in general, generated samples are significantly closer to the reference data of their intended class than to any other category. This behavior confirms that SCAPES successfully captures the distinguishing acoustic features of each environmental texture. Furthermore, the high degree of consistency across VGGish, CLAP, and TexStat representations indicates that this adherence is not merely a reflection of the conditioning space (CLAP), but is also preserved for example in terms of low-level perceptual texture statistics (TexStat). These results suggest that the model synthesizes high-quality textures that are both semantically faithful and stochastically varied, characteristic of natural environmental sounds.

\subsubsection{Long-Term Stability and Autoregressive Drift}\label{subsubsec:long_term_stability}
To assess long-term stability, we conducted a robustness study across all dataset categories. Autoregressive models are prone to error accumulation, which can lead to "drift" toward out-of-distribution states, feedback loops, or collapse into silence or noise. We generated 125-second audio sequences fully autoregressively from a cold start across 100 trials, using semantic context embeddings for conditioning computed from the dataset. CLAP embeddings were then computed over 5-second segments at 30, 60, and 120 seconds, and compared to the corresponding conditioning embeddings.

As shown in Table~\ref{tab:long_term_stability}, a high degree of similarity is maintained over time, indicating stable long-term behavior. These results suggest that SCAPES remains well anchored to the intended semantic manifold, with minimal drift or degradation, and that its conditioning mechanisms operate effectively and in synergy to preserve long-term consistency.

\subsubsection{Qualitative Analysis of Semantic Interpolation}\label{subsubsec:interpolation}
A compelling property of the SCAPES architecture is its ability to perform smooth interpolations between different acoustic classes. By traversing the CLAP conditioning space at inference time, utilizing spherical linear interpolation (Slerp) between two target class embeddings, the model can be steered to generate intermediate textures. Moreover, because SCAPES operates on a continuous latent manifold rather than a discrete codebook, these transitions avoid the "quantization jumps" or sonic artifacts typically associated with token-based models.

Preliminary interpolation results are available on the project page and can be explored through our interactive demo. Qualitative inspection indicates that the model supports smooth transitions between acoustically and semantically related soundscapes. For example, interpolating from a creek to a river gradually increases the perceived intensity and density of flowing water, while transitions from gentle rain to a thunderstorm progressively introduce heavier rainfall, stronger wind, and thunder. More surprisingly, the model also produces plausible intermediate textures between semantically unrelated classes exhibiting similar acoustic characteristics. For instance, an interpolation from keyboard strokes to bonfire crackling gradually dissolves the rhythmic keyboard impulses into stochastic crackling while introducing the low-frequency hum characteristic of a bonfire, ultimately yielding a coherent hybrid environmental texture.

While we acknowledge that the objective evaluation of such hybrid sounds remains an open challenge in generative audio, these results suggest that the learned vector field $F_\theta$ is well-regularized and capable of generalizing to regions of the semantic manifold not explicitly represented in the training data. This capability points toward the potential for SCAPES to serve as a tool for fine-grained creative sound design, allowing users to navigate a continuous spectrum of environmental textures.

\section{Discussion}\label{sec:discussion}
While the trend in generative audio continues toward large-scale networks with billions of parameters trained on vast, proprietary datasets, a counter-movement has emerged prioritizing openness, adaptability, and community-driven agency \cite{musgo, musgu}. SCAPES is positioned within this paradigm of "small-scale" AI, offering a fully open-source pipeline that includes the core architecture, curated data, and pre-trained weights. The primary strength of our model lies in its extreme adaptability; because the framework is trained without manual annotations and is computationally efficient, researchers can curate and train on domain-specific datasets in remarkably short timeframes using consumer-grade hardware. 

\begin{table}[h]
    \centering
    \footnotesize
    \caption{Long-term stability analysis. Values represent the mean cosine similarity ($\pm$ standard deviation, where 1 denotes a perfect match) between the target context embedding and 5-second segments of generation after different periods of time.}
    \setlength{\tabcolsep}{0pt}
    \begin{tabularx}{\columnwidth}{
    >{\raggedright\arraybackslash}X
    >{\centering\arraybackslash}X
    >{\centering\arraybackslash}X
    >{\centering\arraybackslash}X
    }
    \toprule
    & \multicolumn{3}{c}{\textbf{Cosine Similarity} (CLAP)} \\
    \cmidrule(l){2-4}
    \textbf{Source} & \textbf{30\,s} & \textbf{60\,s} & \textbf{120\,s} \\
    \midrule
    Applause  & $0.76 \pm 0.12$ & $0.71 \pm 0.09$ & $0.78 \pm 0.09$ \\
    Bonfire & $0.73 \pm 0.11$ & $0.73 \pm 0.12$ & $0.74 \pm 0.12$ \\
    Bubbles     & $0.86 \pm 0.04$ & $0.89 \pm 0.03$ & $0.88 \pm 0.04$ \\
    Forests    & $0.75 \pm 0.06$ & $0.81 \pm 0.07$ & $0.77 \pm 0.06$ \\
    Helicopter   & $0.93 \pm 0.02$ & $0.93 \pm 0.03$ & $0.86 \pm 0.04$ \\
    Keyboard  & $0.83 \pm 0.06$ & $0.81 \pm 0.06$ & $0.81 \pm 0.07$ \\
    Orchestra & $0.82 \pm 0.07$ & $0.89 \pm 0.03$ & $0.81 \pm 0.09$ \\
    Rain     & $0.87 \pm 0.06$ & $0.91 \pm 0.03$ & $0.91 \pm 0.04$ \\
    Rivers     & $0.80 \pm 0.08$ & $0.83 \pm 0.07$ & $0.81 \pm 0.07$ \\
    Wind  & $0.87 \pm 0.06$ & $0.87 \pm 0.06$ & $0.87 \pm 0.04$ \\
    \bottomrule
    \end{tabularx}

    \label{tab:long_term_stability}
\end{table}

The controllability of SCAPES is another distinguishing factor. The model can operate in several modes: generating indefinite, non-repeating environmental textures from a single semantic embedding; performing "creative continuation" of existing audio; or navigating the semantic manifold via the interpolation strategy detailed in Section~\ref{subsubsec:interpolation}. The success of the latter can be attributed to three key factors. First, the conditioning signal is purposefully "relaxed": since the CLAP embedding is computed from a context window that extends beyond the target atom, the model treats the semantic prompt as a directional guide for a transition rather than a rigid, instantaneous constraint. Second, during training, the model learns to map the distinct "islands" of the sparse CLAP manifold to specific acoustic textures. When traversing the unrepresented space between these clusters, the model must solve a continuity task, synthesizing plausible intermediate states that satisfy both its autoregressive memory and the shifting semantic target. Finally, the use of Optimal Transport Flow Matching \cite{flow-matching} yields a highly regularized, continuous vector field. Because the generative process is deterministically governed by smooth ODE trajectories, gradually shifting the semantic condition produces a smooth evolution in the acoustic output, a property only possible in continuous latent spaces.

Despite its versatility, SCAPES possesses inherent limitations. The model currently struggles with signals requiring strict long-term structural dependencies, such as speech or complex polyphonic music. This is a predictable consequence of the "atomic" design; by focusing on a 10\,Hz control rate and a limited memory buffer, the model lacks the global receptive field necessary to maintain phonemic or harmonic consistency over several seconds. In these contexts, the task of predicting the next 0.1 seconds of audio based purely on local memory and high-level semantics is underspecified. While increasing the memory buffer or incorporating hierarchical conditioning could mitigate these issues, we consider the current architecture to be optimized specifically for the stochastic, textural nature of environmental soundscapes.

\section{Conclusions and Future Work}\label{sec:conclusions}
In this work, we presented SCAPES, a Semantically Conditioned Autoregressive Prior for Environmental Sounds. By combining Continuous Normalizing Flows (CNFs), Flow Matching objectives, and the continuous latent representations of EnCodec, SCAPES provides a robust, lightweight framework for high-fidelity environmental sound synthesis. Our evaluation across 10 distinct acoustic categories demonstrates that a 36-million parameter architecture can successfully produce high-quality stereo textures that exhibit strong semantic adherence and long-term autoregressive stability. Furthermore, by operating on a continuous latent manifold, SCAPES functions as an effective interpolation engine, enabling seamless transitions between disparate semantic states, a capability that remains fundamentally challenging for discrete, token-based architectures.

Several avenues for future research remain open. First, while we utilized CLAP embeddings for semantic conditioning, SCAPES is fundamentally agnostic to the choice of conditioning signal. Future work could explore the integration of low-level acoustic features or human-made annotations. Second, the current model is optimized for a localized receptive field to maintain computational efficiency. Expanding this temporal window through hierarchical attention or sparse transformers could potentially resolve current limitations regarding long-term structural dependencies in speech and music. Third, as the complexity and scale of the training data increase, the scaling laws governing the relationship between model capacity and acoustic diversity warrant further investigation. Finally, while our current inference times are highly competitive, the application of techniques such as rectification and distillation \cite{rectifiedflows} could further accelerate generation, potentially enabling even more complex real-time synthesis applications.

\section{Acknowledgments}
This work has been supported by the project "Cátedra en IA y Música" (TSI-100929-2023-1), funded by the Secretaría de Estado de Digitalización e Inteligencia Artificial, the European Union-Next Generation EU funds and BMAT Music Innovators. We also acknowledge support from NVIDIA Corporation and Meta through academic grant programs. Additionally, we would like to express our sincere gratitude to Daniela Quimis, Clara Charbonnier, Jan Pol Obrador, Marcel Manzano, Omar Hamze, Jordi Fabregat, Eduard Herrera, and Eric Mas, for their valuable insights into extending the model to diverse audio sources, exploring alternative training techniques, and helping shape the design of an intuitive demonstration of SCAPES.

\bibliographystyle{preamble/IEEEtranDAFx}
\bibliography{preamble/bibliography.bib} 


\end{document}

%% file: preamble/auxiliar.tex
\usepackage{etoolbox}

\usepackage{preamble/dafx26v3}

\usepackage{amsmath,amssymb,amsfonts,amsthm}
\usepackage{siunitx}
\usepackage{euscript}
\usepackage[T1]{fontenc}
\usepackage[utf8]{inputenc}
\usepackage{ifpdf}
\usepackage[english]{babel}
\usepackage{caption}
\usepackage{subfig} 
\usepackage{color}
\usepackage{booktabs}
\usepackage{lipsum}
\usepackage{tabularx}
\usepackage{float}


%% file: preamble/authors.tex
\def\papertitle{SCAPES: Semantically Conditioned Autoregressive Prior for Environmental Sounds}
\def\paperauthorA{Esteban Gutiérrez}
\def\paperauthorB{Lonce Wyse}
\def\paperauthorC{Frederic Font}
\def\paperauthorD{Xavier Serra}

\input glyphtounicode
\ninept

\newcounter{numauth}
\newcounter{listcnt}
\newcommand\authcnt[1]{\ifdefined#1 \stepcounter{numauth} \fi}

\newcommand\addauth[1]{
\ifdefined#1 
\stepcounter{listcnt}
\ifnum \value{listcnt}<\value{numauth}
\appto\authorslist{, #1}
\else
\appto\authorslist{~and~#1}
\fi
\fi}
\authcnt{\paperauthorB}
\authcnt{\paperauthorC}
\authcnt{\paperauthorD}
\authcnt{\paperauthorE}
\authcnt{\paperauthorF}
\authcnt{\paperauthorG}
\authcnt{\paperauthorH}
\authcnt{\paperauthorI}
\authcnt{\paperauthorJ}
\def\authorslist{\paperauthorA}
\addauth{\paperauthorB}
\addauth{\paperauthorC}
\addauth{\paperauthorD}
\addauth{\paperauthorE}
\addauth{\paperauthorF}
\addauth{\paperauthorG}
\addauth{\paperauthorH}
\addauth{\paperauthorI}
\addauth{\paperauthorJ}

\usepackage{times}

\newif\ifpdf
\ifx\pdfoutput\relax
\else
   \ifcase\pdfoutput
      \pdffalse
   \else
      \pdftrue
   \fi
\fi

\ifpdf 
  \usepackage[pdftex,
    pdftitle={\papertitle},
    pdfauthor={\authorslist},
    pdfsubject={Proceedings of the 29th International Conference on Digital Audio Effects (DAFx26)},
    colorlinks=false, 
    bookmarksnumbered, 
    pdfstartview=XYZ 
  ]{hyperref}
  \usepackage[pdftex]{graphicx}
\else 
  \usepackage[dvips]{epsfig,graphicx}
  \usepackage[dvips,
    pdftitle={\papertitle},
    pdfauthor={\authorslist},
    pdfsubject={Proceedings of the 29th International Conference on Digital Audio Effects (DAFx26)},
    colorlinks=false, 
    bookmarksnumbered, 
    pdfstartview=XYZ 
  ]{hyperref}
\fi
\usepackage[hypcap=true]{caption}
\title{\papertitle}

\affiliation
{\paperauthorA, \paperauthorB, \paperauthorC\ and \paperauthorD}
{\href{https://www.upf.edu/web/mtg}{Music Technology Group} \\ Universitat Pompeu Fabra\\ Barcelona, Spain\\
{\tt \href{mailto:esteban.gutierrezc@upf.edu}{<esteban.gutierrezc, lonce.wyse, frederic.font, xavier.serra>@upf.edu}}
}